\documentclass{article}
\usepackage{spconf,amsmath,graphicx}
\usepackage{algorithm}  
\usepackage{algpseudocode}  
\usepackage{amsmath}  
\usepackage{multirow}
\usepackage{booktabs}
\usepackage[misc]{ifsym}
\usepackage{hyperref}

\title{A Study of Designing Compact AUDIO-VISUAL WAKE WORD SPOTTING SYSTEM BASED ON ITERATIVE FINE-TUNING IN NEURAL NETWORK PRUNING}

\name{Hengshun Zhou$^1$, Jun Du$^{1,*}$\thanks{\textsuperscript{*}corresponding author}, Chao-Han Huck Yang$^2$, Shifu Xiong$^3$, Chin-Hui Lee$^2$}
\address{$^1$University of Science and Technology of China, Hefei, Anhui, P. R. China \\
$^2$Georgia Institute of Technology, Atlanta, GA. USA \\
$^3$iFlytek Research, Hefei, Anhui, P. R. China \\
\small zhhs@mail.ustc.edu.cn, {\scriptsize \Letter}jundu@ustc.edu.cn, huckiyang@gatech.edu, sfxiong@iflytek.com, chl@ece.gatech.edu}

\begin{document}
\ninept
\maketitle
\begin{abstract}
Audio-only based wake word spotting (WWS) is challenging under noisy conditions due to the environmental interference in signal transmission. In this paper, we investigate on designing a compact audio-visual WWS system by utilizing the visual information to alleviate the degradation. Specifically, in order to use visual information, we first encode the detected lips to fixed-size vectors with MobileNet and concatenate them with acoustic features followed by the fusion network for WWS. However, the audio-visual model based on neural network requires a large footprint and a high computational complexity. To meet the application requirements, we introduce neural network pruning strategy via the lottery ticket hypothesis in an iterative fine-tuning manner (LTH-IF), to the single-modal and multi-modal models, respectively. Tested on our in-house corpus for audio-visual WWS in a home TV scene, the proposed audio-visual system achieves significant performance improvements over the single-modality (audio-only or video-only) system under different noisy conditions. Moreover, LTH-IF pruning can largely reduce the network parameters and computations with no degradation of WWS performance, leading to a potential product solution for the TV wake-up scenario. 
\end{abstract}


\vspace{-0.1cm}

\section{Introduction}
\label{sec:intro}

Wake word spotting (WWS) can be considered as a specific case of keyword spotting (KWS), concerning the identification of pre-defined wake word(s) in utterances, often used for the wake-up of speech-enabled devices, such as “Hey Siri” in iPhone, “Alexa” in Amazon Echo, and “Ok Google” in Google Home~\cite{8683546,9054977,8682682,JoseMSSEV20}, etc. In order to activate the interactions between devices and users, a standby wake word detection module is particularly important~\cite{8683479}.

Traditional approaches in keyword spotting tasks involve the keyword/filler hidden Markov model (HMM)~\cite{115555,9414797}, namely training an HMM for the keyword and a filler HMM for the non-keyword segments, respectively. Recently, deep learning based keyword spotting have attracted much attention. Chen et al. proposed a simple discriminative keyword spotting approach based on deep neural networks which have improved the performance of system~\cite{6854370}. The first attempt to use convolutional neural networks (CNNs) for keyword spotting, by Sainath and Parada~\cite{sainath2015}, was recently improved by jointly integrating deep residual learning and dilated convolutions~\cite{8462688}. Arik et al.~\cite{ArikKCHGFPC17} also applied the convolutional recurrent neural network (CRNN) architecture to single English keyword detection. With the achievements of Transformer~\cite{vaswani2017} in the field of deep learning, several variants of Transformers for wake word detection are explored in~\cite{9414777}. Besides, more efficient networks have been also investigated by leveraging recent advances in differentiable neural architecture search~\cite{9414848}.

Despite the above research progress, KWS is still a challenging task and has attracted the attention of speech researchers. On the one hand, the KWS systems usually perform very well under clean speech conditions. However, the systems suffer from sharp performance degradation under noisy environments. The authors in~\cite{9054538} propose integrating multiple beamformed signals and leveraging the attention mechanism to dynamically tune the model's attention to the reliable input sources to improve the KWS performance under noisy and far-field conditions. The data-efficient solutions are presented in~\cite{9053313} to improve the model robustness in WWS modeling under noisy conditions. A multi-task network that performs KWS and speaker verification (SV) simultaneously is also proposed in~\cite{JungJGK20} to fully utilize the interrelated domain information aiming at performance improvement in challenging conditions. In addition, the authors in~\cite{9465680} have developed a novel tuple-based loss function along with a training strategy for noise-robust keyword spotting.

\begin{figure*}[htp]
    \centering
    \includegraphics[width=17cm]{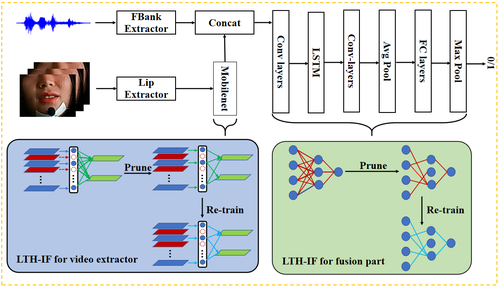}
    \caption{The architecture of proposed audio-visual wake word spotting with neural network pruning using LTH-IF.}
    \label{fig:framework}
\end{figure*}

On the other hand, the WWS system usually runs on smart devices, it’s critical to design the model with a small footprint and low computational power. The application of multi-scale temporal modeling to the small-footprint keyword spotting task has been explored in~\cite{LiWQ20}. The authors also explore the latency and accuracy of KWS models in streaming and non-streaming modes for simplifying model deployment on mobile devices~\cite{RybakovKSVL20}. In~\cite{BlucheG20}, the researchers designed different models and neural architectures for small footprint keyword spotting. Different loss functions for the training of a small-footprint KWS system have also been explored in~\cite{8682534}. Recently, in order to optimize towards memory footprint and execution time, power-consuming audio preprocessing and data transfer steps are eliminated in~\cite{9053395} by directly classifying from raw audio.
 
In order to alleviate the performance degradation of WWS under noisy conditions, in this paper, we investigate an audio-visual WWS system by utilizing the visual information. First, the detected lips are encoded to fixed-size vectors with MobileNet, and concatenated with acoustic features. Next, a neural network pruning strategy, i.e the lottery ticket hypothesis based iterative fine-tuning (LTH-IF) is introduced to the WWS systems. Finally, tested on our in-house corpus for audio-visual WWS in a home TV scene, the proposed audio-visual system achieves significant performance improvements over the single-modality system under different noisy conditions. Moreover, LTH-IF pruning can largely reduce the network parameters and computations with no degradation of WWS performance, leading to a potential product solution for the TV wake-up scenario.

\section{PROPOSED APPROACH}
\label{sec:proposed approach}

\subsection{Audio-Visual Model for Wake Word Spotting}
\label{sec:AV KWS-Net}

 Inspired by the work in~\cite{momeni2020seeing}, we design the proposed audio-visual wake word spotting (WWS) architecture in an end-to-end manner. The main difference from~\cite{momeni2020seeing} lies on that we have the only one wake word, so we do not need to use the phonetic sequence or calculate the similarity matrix. Accordingly we directly adopt the  classification task through the fully connected (FC) layers. The overall flowchart of proposed audio-visual WWS architecture is shown in Fig.~\ref{fig:framework}, which mainly consists of three parts: audio stream, video stream and fusion stream. The details will be elaborated in the following subsections.
 
\subsubsection{Audio Stream}

 For the audio stream, the acoustic features extracted frame by frame are selected as input features. Here, we employ 40-dimensional filter bank (FBank) features normalized by global mean and variance. Given the raw input audio data  ${I}_{\text{A}}$, we can calculate normalized FBank features ${F}_{\text{A}}$ through the FBank extractor $f_{\text{A}}$:
 
 \begin{equation}
    {F}_\text{A} = f_\text{A}({I}_\text{A})
\end{equation}
 
 \subsubsection{Video Stream}
 
 For the video stream, considering practicality and lightweight, we select a combination of MobileNetV2~\cite{sandler2018} as our video embedding extractor. In addition, we replace bidirectional long short-term memory (BLSTM) with LSTM to further reduce the model size and latency. In this study, 13 linear bottlenecks are first adopted as a lip feature extractor $f_\text{V}$. Given the input image ${I}_\text{V}$, the lip feature ${F}_\text{V}$ is calculated through $f_\text{V}$:
  \begin{equation}
    {F}_\text{V} = f_\text{V}({I}_\text{V})
\end{equation}
 The gray scale lip ${F}_\text{V}$ reshaped to 88 $ \times $ 88 is used as the MobileNetV2 input, and the output is an encoded vector by using average pooling. The high-level video embeddings ${E}_\text{V}$ can be obtained by MobileNetV2 $f_\text{Mobile}$:
   \begin{equation}
    {E}_\text{V} = f_\text{Mobile}({F}_\text{V})
\end{equation}
 For more details, please refer to~\cite{sandler2018}. 

 \subsubsection{Fusion Stream}
 
 For the audio-visual fusion stream, a direct concatenation for audio-visual fusion at the encoder is first considered to integrate information from two sources ${F}_\text{A}$ and ${E}_\text{V}$:
   \begin{equation}
    {F}_\text{AV} = [{F}_\text{A},{E}_\text{V}]
\end{equation}
 Then a mixture of convolutional layers, LSTM layers and FC layers are designed to generate the final output as shown in Fig.~\ref{fig:framework}. For the single-modality network, the above audio-visual features ${F}_\text{AV}$ are replaced by audio acoustic features ${F}_\text{A}$ and visual embeddings ${E}_\text{V}$, respectively. In this study, 5 epochs ($E=5$) are selected to train these three WWS models, namely audio-only model, video-only model, and audio-visual model. The final output of these models is compared with the preset threshold after sigmoid operation, `1' indicates that the current sample contains wake word, and `0' indicates the opposite. Given a sample, the model ($G$) outputs a probability $p(y=1|\Theta )$ representing the possibility that the wake word is included, where $\Theta$ represents the model parameter set. The optimisation objective is a binary cross-entropy loss between this prediction and the ground truth label $y^{*}$:

\begin{equation}\label{loss}
\ L_{\text{WWS}}\!=\!-y^{*}\log\ p(y\!=\!1|\Theta )\!-\!(1\!\!-\!y^{*})\ \log(1\!\!-p(y\!=\!1|\Theta )) \
\end{equation}

\begin{algorithm}[htb]  
  \caption{LTH with Iterative Fine-tuning}  
  \label{alg:lth}  
  \begin{algorithmic}[1]  
    \Require  A model, $G_{0}$
    \State Randomly initialize weights $(\Theta _{0})$
    \label{code:label0}  
    \State Initialize model: $G_{0}(\Theta _{0})\rightarrow G_{1}$
    \label{code:label1}  
    \State For $t= 1,\cdots, T: \# \ Pruning \; searching \; iterations$
    \label{code:label2}  
    \State $\quad$ If $t= 1$:
    \label{code:label3} 
    \State $\quad$ $\quad$ For $e= 1,\cdots, E: \# \; Training \; epochs$  
    \label{code:label4}  
    \State $\quad$ $\quad$ $\quad$ $\Theta _{e}\rightarrow \Theta$ : Train $G_{t}$ for its final weights $(\Theta _{t})$
    \label{code:label5}
    \State  $\quad$ Else:
    \label{code:label6} 
    \State  $\quad$ $\quad$ For $e= 1: \# \; Training \; epoch$
    \label{code:label7} 
    \State $\quad$ $\quad$ $\quad$ $\Theta _{e}\rightarrow \Theta$ : Train $G_{t}$ for its final weights $(\Theta _{t})$
    \label{code:label8}
    \State  $\quad$ If $t< T : \# \ LTH \; pruning \; strategy$
    \label{code:label9} 
    \State  $\quad$ $\quad$ Mask$(\Theta _{t})$ to get a pruned graph $G_{p}$ from $G_{t-1}$
    \label{code:label10}
    \State $\quad$ $\quad$ Load weights $\Theta _{p}\in \Theta _{t}$ from $G_{p}$
    \label{code:label11}
    \State $\quad$ $\quad$ Update target model $G_{p}(\Theta _{p})\rightarrow G_{t+1}$
    \label{code:label12}
    \Ensure  
    A well-trained pruned model $G_{T}(\Theta _{T})$ 
  \end{algorithmic}  
\end{algorithm}  

\subsection{Audio-Visual Model Pruning Using LTH-IF}
\label{sec:lth pruning}

The recent ``Lottery Ticket Hypothesis''~\cite{frankle2018lottery} showed an encouraging phenomenon that some subnetworks (winning tickets) could be obtained by pruning the original network through specific methods, and then they can be trained to achieve the performance equal to or better than the untrimmed original model. Although convolutional neural networks have shown to be effective to the small-footprint WWS problem, they still need hundreds of thousands of parameters to achieve good performance~\cite{XuZ20}. Although LTH-based low-complexity neural models have proven competitive prediction performance on several image classification tasks, machine translation~\cite{renda2019comparing,abs210316547} and acoustic scene classification~\cite{abs210701461}, and recently have been supported with some theoretical findings~\cite{abs200200585} related to overparameterization, the effect of LTH on our multimodal task of audio-visual WWS is not unknown. In this study, with a high demand of designing a compact audio-visual WWS model with low-latency for real applications, we investigate on neural network pruning based on LTH with an iterative fine-tuning strategy.

{\bf LTH-IF Algorithm Design:} In Algorithm 1, we detail our approach: First, our WWS model with its original neural architecture $G_{0}$ is initialized with the weights parameters $(\Theta _{0})$. Different from~\cite{frankle2018lottery}, the complete number of iterations ($E=5$) is selected to train the model only before pruning (i.e. $t=1$). At the end of each training phase, a pruning iteration is started if the current iteration $t$ is less than $T$. And the final weights $\Theta _{T}$ are used for the new initialization to fine-tune the model. The LTH pruning searches for a low-complexity model from steps (10) to (13). 

For the audio-only WWS model, we prune the whole network directly based on LTH-IF. However for video-only WWS model, unlike audio-only WWS, it includes an additional lip encoder, and we also prune it jointly with the back-end module based on LTH-IF.

Interestingly, for the audio-visual WWS model, we found that separate pruning for the lip encoder first will yield better performance than pruning the whole model directly at the same degree of pruning. Therefore, for the audio-visual WWS model, we first prune the lip encoder using LTH-IF. Then the lip encoder is fixed and initialized using the weights and mask obtained above. Finally, the model is trained for pruning back-end network of fusion stream based on LTH-IF.

\section{EXPERIMENTS}
\label{sec:experiments}

\subsection{Databases and Implementation Details}
\label{sec:databses and details}

We conduct experiments on an audio-visual dataset collected in smart home TV scenes. There are 250 speakers in total, with a male-to-female ratio of 1:1. The wake word is ``Xiao T Xiao T''. The speakers in the training set, development set and test set do not overlap, which are 210, 20 and 20 respectively. For the training set, in addition to the original positive and negative samples, we also added several types of noises for data augmentation. Our final training set includes 50 hours of positive samples and 50 hours of negative samples, respectively. For the development/test set, in order to facilitate comparison, we only add noise with three signal-to-noise ratios (i.e. -5dB, 0dB, 5dB), including 2 hours of positive samples and 2 hours of negative samples, respectively. Each positive sample contains only one wake word. The duration of each sample is 1.3 seconds.

We use false reject rate (FRR) and false alarm rate (FAR) on test set as the criterion of the WWS performance. Suppose the test set consists of $N_{\text{wake}}$ examples with wake word and $N_{\text{non-wake}}$ examples without wake word, FRR and FAR are defined as follows:

\begin{equation}\label{frr}
\ \text{FRR}=\frac{N_{\text{FR}}}{N_{\text{wake}}} \
\end{equation}

\begin{equation}\label{far}
\ \text{FAR}=\frac{N_{\text{FA}}}{N_{\text{non-wake}}} \
\end{equation}

\noindent where $N_{\text{FR}}$ denotes the number of examples including wake-up word but the WWS system gives a negative decision. $N_{\text{FA}}$ is the number of examples without wake-up word but the WWS system gives a positive decision. 

We employ pytorch to train all models and minimize the loss function using the Adam optimization method. The batch size is 64 for audio-only WWS system and 16 for video-only and audio-visual systems. The learning rates are set to 0.0001, 0.0002, 0.0002 for audio-only, video-only and audio-visual systems respectively.

\subsection{Results on Audio-Visual Wake Word Spotting}
\label{sec:results av-kws}

First we evaluate the performance of the single-modal systems. We train the audio-only and the video-only system respectively, corresponding to the audio stream and video stream of the audio-visual model shown in Fig.~\ref{fig:framework}. Our results are presented in Table~\ref{tab:avkws results}. We can observe that the better results were achieved by audio-only system compared with video-only especially in less-noisy environments. In low-SNR adverse acoustic environments, video-only system achieves better FAR performance, which indicates potential of audio-visual system that integrates the advantages of audio-only and video-only systems. Moreover, a direct concatenation for audio-visual fusion at the encoder part is further implemented, which yielded remarkable improvements compared with the single-modal system under various signal-to-noise ratios. For example, the performance gap for -5dB is 4.78\% of FAR between audio-only model and audio-visual model.

\begin{table}[th]
    \caption{Test set performance comparison of different systems.}
    \centering
    \label{tab:avkws results}
    \begin{tabular}{ccccc}
        \toprule
        \multirow{2}{*}{Modality} & \multirow{2}{*}{1-FRR (\%)} & \multicolumn{3}{c}{FAR (\%)}\\
        &   & -5dB & 0dB & 5dB\\
        \midrule
        Audio & 98.78 & 8.03 & 2.95 & 1.60 \\
        Video & 98.78 & 6.92 & 6.92 & 6.92 \\
        Audio-visual & 98.78 & 3.25 & 1.06 & 0.56 \\
        \bottomrule
    \end{tabular}
\end{table}

\begin{table}[th]
 \caption{Parameter statistics of different systems.}
 \label{tab:parameter statistics}
 \centering
 \setlength{\tabcolsep}{3mm}{
 \begin{tabular}{ccccc}
  \toprule
  Network & Param. (M) & FLOPs (M) \\
  \midrule
  Audio & 0.35 & 11.29 \\
  Lip Encoder & 0.39 & 642.56 \\
  Video & 1.19 & 651.68\\
  Audio-Visual & 1.55 & 656.48 \\
  \bottomrule
 \end{tabular}}
\end{table}

The statistics of the parameters and FLOPs of these three models are shown in Table~\ref{tab:parameter statistics}. We can observe that after adding video modality, the number of parameters and FLOPs of the audio-visual WWS model is greatly increased. In particular, the parameters and FLOPs of lip encoder exceed those of audio-only network, which also promotes us to explore more effective pruning methods.

\begin{table}[ht]
    \caption{Test set performance of single-modalities and multi-modality systems under different pruning degrees. [1-FRR : 98.78\%]}
    \label{tab:single-modal pruning}
    \centering
    \setlength{\tabcolsep}{1.1mm}
    \begin{tabular}{cccccc}
        \toprule
        \multirow{2}{*}{Modality} & \multirow{2}{*}{Method} & \multicolumn{3}{c}{FAR (\%)} & \multirow{2}{*}{Pruned (\%)}\\
        &   & -5dB & 0dB & 5dB \\
        \midrule
        \multirow{3}{*}{Audio} & No Pruning & 8.03 & 2.95 & 1.60 & 0.00 \\
         & LTH~\cite{frankle2018lottery} & 34.00 & 23.67 & 18.3 & 71.90 \\
         & LTH-IF  & 7.71 & 2.26 & 1.15 & 80.21 \\
        \midrule
        \multirow{3}{*}{Video} & No Pruning & 6.92 & 6.92 & 6.92 & 0.00 \\
         & LTH~\cite{frankle2018lottery} & 16.11 & 16.11 & 16.11 & 29.86 \\
         & LTH-IF  & 6.89 & 6.89 & 6.89 & 55.65 \\
        \midrule
        \multirow{2}{*}{Audio-Visual} & No Pruning & 3.25 & 1.06 & 0.56 & 0.00 \\
         & LTH-IF & 2.29 & 0.84 & 0.53 & 42.52 \\
        \bottomrule
    \end{tabular}
\end{table}

\vspace{-0.5cm}

\subsection{Results on LTH-IF Pruning}
\label{sec:results lth}

We first implemented LTH-based pruning on single-modal systems described in~\cite{frankle2018lottery} with the results shown in Table~\ref{tab:single-modal pruning}. When LTH-based pruning in~\cite{frankle2018lottery} is firstly applied to the single-modal models, the performance degrades rapidly in both audio and video modalities. 
Compared to original unpruned model, using LTH with iterative fine-tuning (LTH-IF) achieves better performance especially in the audio modality even though over 80\% model parameters are pruned, which demonstrates the effectiveness of the iterative fine-tuning strategy.

\begin{figure}[htp]
    \centering
    \includegraphics[width=8.5cm]{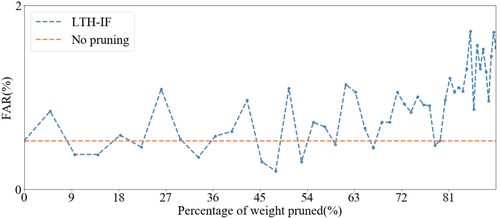}
    \caption{Test set performance on audio-visual WWS model with the iterative pruning during the training process.}
    \label{fig:pruning_curve1}
\end{figure}

Based on these positive results, we next apply LTH-IF pruning to the audio-visual WWS model, and the results are shown in last row of Table~\ref{tab:single-modal pruning}. 
According to Table~\ref{tab:parameter statistics}, most of the parameters and FLOPs in audio-visual system come from the lip encoder. Thus we design an experiment by only applying LTH-IF pruning for the lip encoder. The pruned model achieves better performance than the audio-visual unpruned model after pruning about 80\% of the parameters in lip encoder. Fig.~\ref{fig:pruning_curve1} shows the results comparison of applying LTH-IF pruning (blue) to the audio-visual WWS model and original unpruned model (red). 
According to the experimental results above, the system performance gradually deteriorates after about 80\% of the model parameters are pruned. Finally, we initialize the mask in LTH-IF according to the lip encoder result, and then apply LTH-IF to the whole audio-visual model for pruning. According to Table~\ref{tab:single-modal pruning}, after pruning 42.52\% of the parameters, we achieve better performance compared to the model without pruning under all three signal-to-noise ratios. 

We randomly select a specific iteration (e.g. $ t = 20$) and list the pruned parameters of different layers of the model which is shown in Table~\ref{tab:pruning analysis}. It can be observed that for both single-modality and multi-modality systems, the pruning proportions of different types of layers are similar. 

However, during the pruning process, we observe that the performance of video-only system is often unstable (degradation with potentially important nodes pruned) and more sensitive to the pruning proportion setting compared with the audio-visual system. So the audio-visual system seems more robust to pruning, which might be explained by that the audio-visual fusion leads to the selection of more suitable nodes without being pruned.

\begin{table}[th]
 \caption{The comparison of pruned parameters after the same iterations for single-modalities and multi-modality systems. [$t$=20]}
 \label{tab:pruning analysis}
 \centering
 \setlength{\tabcolsep}{1.5mm}{
 \begin{tabular}{cccc}
  \toprule
  Modality & Conv layers(\%) & LSTM(\%) & FC layers(\%) \\
  \midrule
  Audio & 65.54 & 65.24 & 65.72 \\
  Video & 65.54 & 65.86 & 65.73 \\
  Audio-Visual & 65.53 & 65.89 & 65.70  \\ 
  \bottomrule
 \end{tabular}}
\end{table}

\vspace{-0.5cm}

\section{CONCLUSION}
\label{sec:conclusion}

In this paper, we investigate on designing a compact audio-visual WWS system under noisy conditions by utilizing video information to alleviate the performance degradation. Tested on our in-house corpus for audio-visual WWS in a home TV scene, the proposed audio-visual system achieves significant performance improvements over the single-modality system under different noisy conditions. Furthermore, a neural network pruning strategy via LTH in an iterative fine-tuning manner is introduced which can largely reduce the network parameters and computations with no degradation of WWS performance.

\section{Acknowledgement}
This work was supported in part by the National Natural Science Foundation of China under Grant No. 62171427, and Alibaba Group.

\vfill\pagebreak

\bibliographystyle{IEEEbib}
\bibliography{strings,refs}

\end{document}